\documentclass[prb,twocolumn,showpacs,amsmath,amssymb]{revtex4}

\usepackage{graphicx}
\usepackage{dcolumn}
\usepackage{bm}

\begin{document}

\title{The effect of nonmagnetic impurities on the local density of states in 
$s$-wave superconductors}%

\author{P. Miranovi\' c}
\author{M. Ichioka}
\author{K. Machida}
\affiliation{Department of Physics, 
Okayama University, 700-8530 Okayama, Japan}

\date{\today}

\begin{abstract}
We study the effect of nonmagnetic impurities on the local density of 
states (LDOS) in $s$-wave superconductors. The quasiclassical equations of
superconductivity are solved selfconsistently to show how LDOS 
evolves with impurity concentration. The spatially averaged zero-energy 
LDOS is a linear function of magnetic induction in low 
fields, $\overline{N(E=0)}=cB/H_{c2}$, for all impurity
concentration. The constant of proportionality $c$ depends 
weakly on the electron mean free path. We present numerical data for
differential conductance and spatial profile of zero-energy LDOS
which can help in estimating the mean free path through the 
LDOS measurement. 
\end{abstract}


\maketitle

\section{Introduction}

Since Hess and co-workers\cite{hess} succeeded in measuring 
the LDOS in the superconducting NbSe$_2$
there have been many reports and theoretical studies on the
electronic structure of the superconductor in the mixed state. 
The novel experimental technique introduced, scanning tunneling
spectroscopy, enables one to measure differential conductivity 
(DC) $\sigma(\bm r,V)$ at various positions $\bm r$ and 
bias-voltages $V$. DC is closely related to
LDOS of the superconductor ($k_B=1$), 
\begin{equation}
\frac{\sigma(\bm r,V)}{\sigma_N}=\int\limits_{-\infty}^\infty
\frac{dE}{4T}
\frac{N(\bm r,E)}{N_0\cosh^2\left(
\dfrac{E+eV}{2T}
\right)}.
\label{dc}
\end{equation}
where $\sigma_N$ is DC in the normal state, $e$ is electron charge,
$N(\bm r,E)$ is LDOS at the position $\bm r$ and energy $E$ relative to 
the Fermi level, and $N_0$ is DOS at the Fermi level in the normal state.
Only in the limit of zero temperature $T\rightarrow 0$ DC and LDOS are 
proportional: $\sigma(\bm r,V)/\sigma_N=N(\bm r,|e|V)/N_0$. At finite 
temperature, DC is actually thermally broadened LDOS. 

It is clear that at low temperatures DC should follow the spatial structure 
of LDOS. Two prominent features should be mentioned.  DC measured at 
the vortex center revealed a peak at the Fermi level (zero-bias peak) that 
well exceeds  $\sigma_N$. This indicates that vortex core can not be
viewed as being ``normal'' at least in clean superconductors.
The zero-bias peak in  DC originates from the 
zero energy peak of LDOS at the vortex center, which 
is due to the low lying bound states inside the vortex core. 
The other remarkable feature revealed 
in Ref. \onlinecite{hess} is a star-shaped DC around the vortex core 
measured at fixed bias-voltage, with star orientation depending on the 
bias-voltage value. The six-fold structure of DC in NbSe$_2$ is coming 
either from the  effect of the hexagonal vortex lattice, anisotropic 
$s$-wave pairing or anisotropic Fermi surface, and most probably 
it is coming from each effect simultaneously. Again, the star-shaped DC 
originates from the star shaped LDOS in the vortex lattice.

However, the measured DC does not follow the sharp features of the
corresponding, theoretically calculated, LDOS even if the experiment is 
performed at very low temperature.\cite{hess1} The height and width of the
zero-bias peak was found to be sample dependent indicating impurities
as a plausible explanation for the discrepancy. Indeed, impurities are 
inevitably  present in superconducting samples on which the experiments 
are performed. Therefore, it is important to quantitatively study how LDOS
is changing with impurity concentration. This is the purpose of 
our paper. 

There is another one topic that we analyze in this paper:
the effect of impurities on the specific heat field dependence.
In $s$-wave superconductors low energy quasiparticles are trapped inside
the vortex core. Therefore, zero-energy LDOS, spatially averaged, is 
proportional to the number of vortexes:  $\overline{N(E=0)}
\propto N_0\xi^2 B$, with $B$ being the magnetic induction and $\xi$ the size
of the vortex core. This translates into the linear field dependence 
of low temperature specific heat given by $C_s/T=2\pi^2 
\overline{N(E=0)}/3$. The nonlinearity in low field $C_s(B)$ curve
should be related to the  gap anisotropy. In the case of anisotropic 
$s$-wave pairing, addition of nonmagnetic impurities can  
smear out the gap anisotropy, which can be tracked by examining 
$C_s(B)$ curves. This kind of measurement has been performed on
Nb$_{1-x}$Ta$_x$Se$_2$\cite{nohara} and
Y(Ni$_{1-x}$Pt$_x$)$_2$B$_2$C.\cite{nohara,lipp}
The intention was to make the gap isotropic by adding
impurities, but with the price to have rather dirty 
$s$-wave superconductor. The effect of impurities on LDOS notwithstanding 
it would be of value to study how field dependence of spatially averaged 
LDOS evolves with impurity concentration in the most simple case
of $s$-wave superconductors.

So far, the only systematic experimental study of the effect of disorder
on LDOS is by Renner {\it et al}.\cite{renner} In particular they 
measured the zero-bias DC at the vortex center in the alloy system 
Nb$_{1-x}$Ta$_x$Se$_2$. Substitution of Nb by Ta leads to systematic decrease
of the electron mean free path. On the other hand small changes in the 
electronic spectrum is expected since Nb and Ta are isoelectronic and with
similar atomic radii. Zero-energy DC is found to be very sensitive to the 
impurity concentration. It gradually disappears and for $x=0.2$ the 
zero-energy LDOS in the vortex center is the same as that of normal phase 
$N_0$. It was even proposed that DC spectra can serve as a measure of 
quasiparticle scattering time.

In this paper the problem of LDOS in presence of nonmagnetic impurities will
be studied within the quasiclassical equations of superconductivity. 
Quasiclassical approximation is adequate in superconductors where
coherence length $\xi$ is much larger than the atomic length $k_F^{-1}$. 
LDOS is studied within the quasiclassical approximation 
by Ullah {\it et al.} \cite{ullah} and Klein \cite{kleinsingle}
for the case of isolated vortex in the isotropic $s$-wave superconductor. 
The full selfconsistent analysis of LDOS in the case of vortex lattice is
performed by Ichioka {\it et al}.\cite{ichioka}
All these studies assume clean superconductor without impurities.
Dirtiness of the superconductor is only roughly estimated by 
Klein.\cite{kleinsingle} As for the single vortex case, the
effect of impurities was studied in Refs. \onlinecite{eschrig,kato}.
The effect of impurities on DOS in extremely high field, where the
Landau level quantization of the electronic energies should be taken 
into account, is studied by Dukan and Te\v sanovi\' c.\cite{dukan}
Those phenomena are beyond the scope of this text. As for the
study of the extreme case, dirty superconductor, the reader is refered to
Refs. \onlinecite{wats,golubov}.

The paper is organized as follows. In  section \ref{method}
the method of solving Eilenberger equations is presented. 
The readers  not interested in technical details may skip 
that section. In section \ref{ldosanddc} spatial and energy
dependence of LDOS and DC for various impurity concentration
is shown. In section \ref{ldosandsh} the effect of
impurities on the specific heat field dependence is discussed. 

\section{Method of solution}
\label{method}

There are various methods to solve Eilenberger equations for 
the vortex lattice. The main problem in the numerical procedure
is that the initial conditions for the differential
equations are unknown. One way to overcome that problem is to use special,
divergent, gauge  in which Green's functions are periodic, and solve equations
in Fourier space (periodic boundary condition).\cite{lowk}
The other method is based on the fact that during the integration process 
Green's functions exponentially grow (explode). Fortunately,
the exponentially growing unphysical solutions can be manipulated to 
form the physical one. This is the essence of the  so-called 
``explosion method''.\cite{thuneberg,klein,ichioka} 
Here we will use a different approach.
It is interesting that if one parameterizes quasiclassical Eqs.
in the form of  Riccati's differential equation, then during the 
numerical integration the physical solution is stabilized regardless of 
the initial condition. Here we will give more details.

For the $s$-wave superconductor in presence of nonmagnetic impurities
Eilenberger equations are
\begin{equation}
\left[\omega+\mathbf u\left(\mathbf\nabla+
\mathrm i\mathbf A\right)\right]f=\Psi g+Fg-Gf,
\end{equation}
\begin{equation}
\left[\omega-\mathbf u\left(\mathbf\nabla-
\mathrm i\mathbf A\right)\right]f^\dagger=\Psi^* g+F^*g-Gf^\dagger.
\end{equation}
These are supplemented by the self-consistency equations for the
gap function $\Psi$ and vector-potential $\mathbf A$
\begin{equation}
\Psi\ln{t}=2t\sum\limits_{\omega>0}\left[
\left<f\right>-\frac{\Psi}{\omega}
\right],
\label{selfdelta}
\end{equation}
\begin{equation}
\mathbf\nabla\times\mathbf\nabla\times\mathbf A=-
\frac{2t}{\widetilde{\kappa}^2}\;\mathrm{I}\mathrm{m}
\sum\limits_{\omega>0}\left< \mathbf u g\right>,
\label{selfa}
\end{equation}
as well as for the impurity potentials
\begin{equation}
F=\frac{1}{\tau}\left<f\right>,\qquad
G=\frac{1}{\tau}\left<g\right>.
\label{selfimp}
\end{equation}
Born approximation is assumed in treating scattering on impurity.
For convenience, equations are written in following dimensionless units:
order parameter is measured in units $\pi T_{c}$, length in units
$R_0=v/(2\pi T_{c})$, $v$ is Fermi velocity,
magnetic field in units $H_0=\Phi_0/2\pi R_0^2$,
where $\Phi_0$ is flux quantum.
Vector-potential is in units $A_0=\Phi_0/2\pi R_0$, energy in units $E_0=
(\pi T_{c})^2N_0R_0^3$. Scattering time $\tau$ is in units
$1/(2\pi T_c)$. It can be expressed via electron mean free path
$l$: $\tau=l/0.882\xi_0$, with $\xi_0$ being BCS coherence length.
Eilenberger parameter $\tilde{\kappa}$ is the only material constant
that enters the equations
\begin{equation}
\widetilde{\kappa}^{-2}=2\pi N_0\left(
\frac{\pi}{\Phi_0}
\right)^2
\frac{v^4}{(\pi T_{c})^2}\,.
\end{equation}
It is related to GL parameter $\kappa$ via $
\widetilde{\kappa}^2=(7\zeta(3)/18)\kappa^2$ 
in 3D case and
\begin{equation}
\tilde{\kappa}^2=\frac{7\zeta(3)}{8}\kappa^2\,.
\end{equation}
in 2D case. Here $\zeta$ is Riemmann's zeta function.
$\omega=t(2n+1)$ is Matsubara frequency with integer $n$, 
$t=T/T_c$ is reduced temperature, $\mathbf u$ is unit vector directed 
along Fermi velocity. Eilenberger Green's functions $f$, $f^\dagger$ and 
$g$ are normalized so that $g=\sqrt{1-ff^\dagger}$. 
Fermi surface is assumed to be isotropic and two-dimensional.\cite{remark}
Average over the isotropic cylindrical Fermi surface reduces to 
$\left<\cdot\cdot\cdot\right>=
(1/2\pi)\int\cdot\cdot\cdot\; d\varphi$, average over polar angle $\varphi$.

The quantity of our interest, LDOS as a function of position $\bm r$ and
quasiparticle excitation energy $E$, is defined as
\begin{equation}
N(\bm r,E)=N_0\left<
{\rm Re}\;g(\bm r,\bm u,\omega\longrightarrow \delta-iE)
\right>,
\end{equation}
where Eilenberger function $g$ describes normal excitations and
$\delta$ is a small number.
It is very convenient to 
introduce auxiliary functions $a$ and $b$ through the
following transformation \cite{schopohl,schopohlandmaki}

\begin{equation}
f=\frac{2a}{1+ab},\quad
f^\dagger=\frac{2b}{1+ab},\quad
g=\frac{1-ab}{1+ab}.
\end{equation}
Equations for auxiliary functions $a$ and $b$ are decoupled
and have the form of Riccati's differential equation 

\begin{equation}
\mathbf u\mathbf\nabla a=-\left(\omega+G+
\mathrm i\mathbf u\mathbf A\right)a+\frac{\Psi+F}{2}
-\frac{a^2}{2}\left(\Psi^*+F^*\right),
\label{ricc1}
\end{equation}
\begin{equation}
\mathbf u\mathbf\nabla b=\left(\omega+G+
\mathrm i\mathbf u\mathbf A\right)b-\frac{\Psi^*+F^*}{2}
+\frac{b^2}{2}\left(\Psi+F\right).
\label{ricc2}
\end{equation}
Auxiliary functions $a(\bm r,\bm u,\omega)$ and 
$b(\bm r,\bm u,\omega)$ are not independent.
Once we solve the Eq. (\ref{ricc1}), function $b$ can
be readily calculated:
\begin{equation}
b(\bm r,\bm u,\omega)=-a^*(-\bm r,\bm u,\omega).
\end{equation}
In the coordinate system $(\rho,\eta)$, where Fermi velocity direction 
$\bm u$ coincides with $\rho$-axis, 
\begin{eqnarray}
\rho=x\cos\phi+y\sin\phi,\nonumber\\
\eta=y\cos\phi-x\sin\phi,
\end{eqnarray}
Eq. (\ref{ricc1}) reduces to
\begin{equation}
\frac{\partial a}{\partial \rho}=-\left(\omega+G+
\mathrm i\mathbf u\mathbf A\right)a+\frac{\Psi'}{2}
-\frac{a^2\Psi'^*}{2},
\label{riccfinal}
\end{equation}
where $\Psi'=\Psi+F$.  Integrating along the direction $\rho$ from 
$\rho'-\rho_\infty$ to the desired point $\rho'$, the physical solution 
$a_+$ is stabilized. Note that integrating in the opposite direction,
toward decreasing $\rho$, one will get solution $a_-=-1/a_+$. How long 
integration path $\rho_\infty$ should be taken  depends on
$\omega$ (is it real or complex) and on impurity concentration.

Vector-potential is written as
\begin{equation}
\bm A(r)=\frac{\bm B\times\bm r}{2}+\bm A'(\bm r),
\end{equation}
where $B$ is magnetic induction, and $\bm A'$ is periodic
with $\bm\nabla\cdot\bm A'=0$. Therefore, the selfconsistent equation
for vector-potential can be written as
\begin{equation}
\nabla^2\bm A'=\frac{2t}{\tilde{\kappa}^2}{\rm Im}
\sum\limits_{\omega>0}\left <\bm u g\right>.
\end{equation}
It can be easily solved in the Fourier space.

Auxiliary function $a$ has the same symmetry properties as 
Eilenberger function $f$, which are described in Ref. 
\onlinecite{ichioka}. The equilibrium vortex lattice structure 
is assumed to be hexagonal. Therefore, it is sufficient to solve equation 
(\ref{riccfinal}) in the whole vortex lattice cell and only for 
velocity directions $0<\varphi<\pi/6$. With the help of symmetry
properties, $a(\bm r,\bm u,\omega)$ can be obtained for all
velocity directions.

\subsection{Iteration procedure}

Iterative procedure for solving Eq. (\ref{riccfinal}) is the following.
We start from some potentials $\Psi(\bm r)$,
$\bm A'(\bm r)$, $F$ and $G$. It is usual to start from
the Abrikosov solution for $\Psi(\bm r)$, $\bm A'(\bm r)=0$, and
local values of impurity potentials:
\begin{equation}
F=\frac{1}{\tau}\frac{\Psi(\bm r)}{
\sqrt{\omega^2+|\Psi(\bm r)|^2}},\qquad
G=\frac{1}{\tau}\frac{\omega}{
\sqrt{\omega^2+|\Psi(\bm r)|^2}}.
\end{equation}
After solving the Eq. (\ref{riccfinal}) the new values of
potentials are obtained from the self-consistency eqs.
(\ref{selfdelta}), (\ref{selfa}), and (\ref{selfimp}), and the
new potentials we plug again into the Eq. (\ref{riccfinal}) and solve it. 
This iterative
procedure is repeated until the selfconsistency is achieved.
The maximum frequency $\omega_{cut}=
t(2N_{cut}+1)$ should be chosen so the result does not
depend on the number of Matsubara frequencies.
On the other hand the number of iteration cycles needed
to stabilize pair potential increase with the $N_{cut}$.  
We followed Klein \cite{klein} and choose  
$\omega_{cut}=20\pi T_c$ (in physical units)
as appropriate for various temperatures. This gives the number
of Matsubara frequencies
\begin{equation}
N_{cut}\approx {\rm Int}\left(\frac{10}{t}\right).
\end{equation}
Fortunately it is not necessary to solve Eq. (\ref{riccfinal})
for all $\omega$. For high frequencies the solution can be well
approximated by:

\begin{equation}
a\approx\frac{1}{2}\left(
\frac{1}{\omega'}-\frac{\bm u\bm\Pi}{\omega'^2}+
\frac{(\bm u\bm\Pi)^2}{\omega'^3}
\right)\left(\Psi+F\right),
\label{approx}
\end{equation}
where $\omega'=\omega+1/\tau$ and $\bm\Pi=\bm\nabla+i\bm A$.
For all $n>N_{cut}/2$ we use the equation (\ref{approx}).
Solution is quasi-periodic. Translation by $\bm R_{nm}=
n\bm r_1+m\bm r_2$ will amount in phase factor 
$\exp(i\chi(\bm r,\bm R_{nm}))$
\begin{equation}
a(\bm r+\bm R_{nm},\bm v,\omega)=
a(\bm r,\bm v,\omega)e^{i\chi(\bm r,\bm R_{nm})},
\end{equation} 
where $\bm r_1$ and $\bm r_2$ are primitive vectors of vortex lattice,
$n$, $m$ are integers, and 
\begin{equation}
\chi=\pi\left[
\frac{mx}{a_0}-\frac{y(n+m\cos\beta)}{a_0\sin\beta}
+nm+n-m
\right].
\end{equation}
The angle between primitive vectors is denoted as $\beta$ ($\beta=\pi/3$
in our case).
Once the selfconsistent potentials $\Psi(\bm r)$ and $\bm A(\bm r)$ are 
calculated, the Eilenberger Eqs. are solved again but this time for
$\omega=\delta-iE$ where $\delta$ is small number and $E$ is quasiparticle 
energy. Note that in the presence of impurities the Eq.
(\ref{riccfinal}) has to be solved 
selfconsistently with respect to impurity potentials $F$ and $G$.
As for the choice of $\delta$ one should be very careful. It was already 
noted\cite{klein1,dahm} that density of states $N(E=0)$ is very sensitive
to the absolute value of $\delta$. Finite $\delta$ has roughly the effect
of impurities and suppresses the peak in DOS at the vortex center.
For small values of  $\delta$, $N(E=0)$ is spiked at the vortex centers
and very fine mesh is needed to evaluate average LDOS. We find
that $\delta=0.001$ suffice for our calculation.

\section{Local density of states and differential conductance}
\label{ldosanddc}
The physics of the vortex core in the clean limit is very different
from the physics of the vortex core in dirty superconductors.
Properties of the vortex core are governed by Andreev bound states 
in the clean limit, while in the dirty limit properties of the core are 
governed by normal electrons.\cite{rainer} To understand the
role of impurities we briefly explain the formation of bound states.
Andreev scattering from the pair potential (order parameter) inside the 
vortex core converse electron-like excitation into hole-like excitation and 
vice versa. States inside the core are coherent superposition of particle 
and hole states. At certain energies the coherent superposition of particle 
and hole states is constructive and the bound state is formed. 
The lowest bound state has the energy $E=\Delta/k_F\xi$. 
In the quasiclassical limit $k_F\xi\gg 1$, 
the lowest bound state energy is pushed to zero.
Zero-energy bound state inside the vortex core will manifest as 
a peak in zero-energy LDOS at the vortex center. Scattering on impurities 
will randomize the motion of electron, and the coherency is lost. Thus, 
the impurities will smear out the sharp structure of LDOS. To illustrate 
this we focus on the spatial structure of zero-energy LDOS $N(\bm r,E=0)$.

\begin{figure}[t]
\includegraphics[angle=0,scale=0.35]{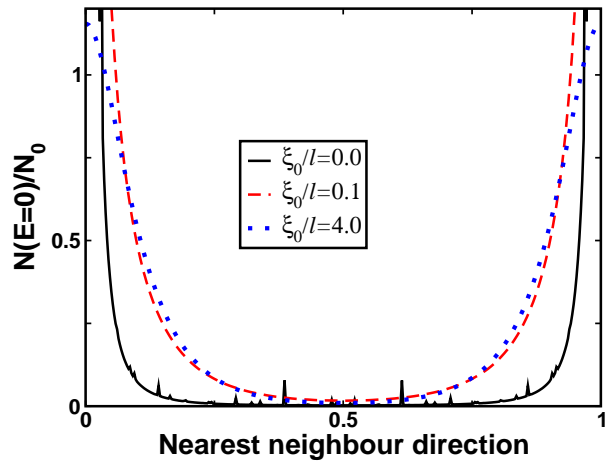}
\caption{Spatial variation of zero energy DOS along 
the nearest neighbor vortex direction. Full line corresponds to the
clean limit and dashed lines correspond to the superconductors with
$\xi_0/\ell=0.1$ and $\xi_0/\ell=4.0$. The calculation is performed at 
approximately the same relative field $B=0.1H_{c2}$.}
\label{nn}
\end{figure}

In Fig. \ref{nn} spatial variation of LDOS along the line 
connecting two nearest neighbor vortexes is shown.
Data for a clean superconductor ($\xi_0/\ell=0.0$), for
a relatively large mean free path $\xi_0/\ell=0.1$ and for impure
case $\xi_0/\ell=4.0$ are presented. Distance between vortexes is 
normalized so that $0$ and $1$ on the abscissa are position of vortexes. 
To remind the reader again, in the clean limit the height and width of the 
LDOS peak depend on the small parameter $\delta$, which measures how far we 
are from the pole of the Green function $g$. In this sense height and width 
of the peak in the clean limit are arbitrary.

\begin{figure}[h]
\hbox{
\includegraphics[angle=0,scale=0.5]{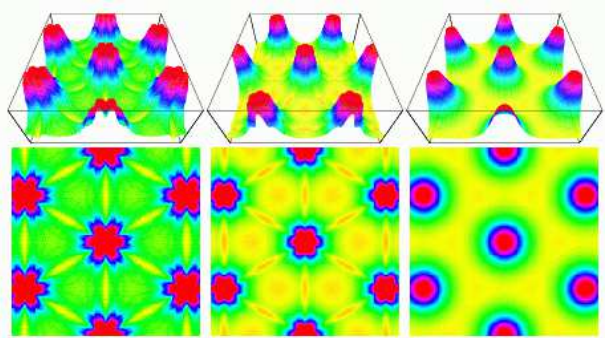}
}
\caption{ZEDOS within the vortex lattice for superconductors
with $\xi_0/\ell=0.0$, $\xi_0/\ell=0.1$ and $\xi_0/\ell=4.0$
(in order from left to right). Only data points $N(E=0)/N_0<1$ are 
presented. Small parameter $\delta=0.03$ is used for clean limit 
data to clarify the spatial distribution.
}
\label{clean}
\end{figure}


At the vortex core zero energy DOS (ZEDOS) in the clean limit highly 
exceeds the normal state value $N_0$. This was in the beginning at odds 
with generally accepted naive picture of vortex core as being ``normal''. 
Analyzing the ZEDOS in the impure case, 
it is clear that coherency is crucial in forming the main peak at the vortex.
In the dirty limit $\xi_0/\ell\rightarrow\infty$ ZEDOS within the vortex core 
approaches normal state value  $N_0$, and only in this limit one can view 
vortex core as being ``normal''. Even a small impurity concentration has a 
great impact on ZEDOS profile.  The comparison of the ideal case of a clean 
superconductor $\xi_0/\ell=0$ with rather pure superconductor 
$\xi_0/\ell=0.1$ reveals a change of the vortex core size by a factor 2. 
The change of the vortex core size is compensated by the reduction of the 
peak height, so the ZEDOS averaged over vortex lattice cell is approximately 
the same in all cases.

\begin{figure}[t]
\includegraphics[angle=0,scale=0.45]{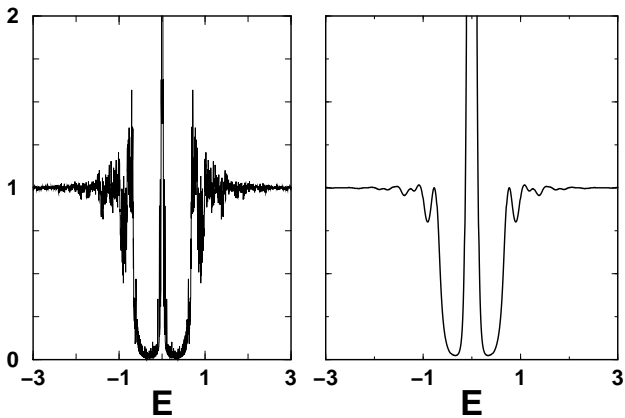}
\caption{a) LDOS $N(E,\bm r=0)/N_0$ at the vortex center as a function of 
excitation energy $E$ (in units $\pi T_c$). b)
DC $\sigma(E)/\sigma_N$ at the vortex center at $T=0.1T_c$.
Both data are for superconductor in the clean limit.}
\label{doseclean}
\end{figure}

It is instructive to see how the spatial structure of ZEDOS within the
vortex lattice evolves by adding impurities. In the clean limit
ZEDOS around the single vortex is cylindrically symmetric.
As soon as vortex lattice is formed cylindrically symmetric ZEDOS 
transforms into the star-shaped structure within the hexagonal vortex 
lattice.\cite{ichioka} This is presented in Fig. \ref{clean}. 
The effect of vortex lattice notwithstanding, the other effects 
such as the anisotropy of the pairing function\cite{hayashi} and the
anisotropy of the Fermi surface in hexagonal crystal can also contribute 
to the specific star-shaped structure of ZEDOS. 
By reducing the mean free path, star-shaped structure gradually disappears
and is completely absent in the dirty limit even at relatively high fields.
This indicates that periodicity of the order parameter is not
the key element to explain the structure of $N(\bm r,0)$ in Fig. \ref{clean}.
Only coherent superposition of electron and hole states in the periodic
vortex lattice can account for the star-shaped ZEDOS.

In Fig. \ref{doseclean}a) LDOS at the vortex center is plotted as a function
of quasiparticle excitation energy $E$ (in units $\pi T_c$) for the clean
case. LDOS oscillates with energy, the result previously reported in Ref. 
\onlinecite{pottinger}. This phenomenon has the same origin as oscillation of 
DOS in superconducting-normal proximity systems: \cite{degennes,tomasch,rowell}
interference of quasiparticles reflected at the superconducting-normal
barrier. The mixed state can be viewed as periodically arranged infinite 
number of ``normal''\--superconducting boundaries. Here the vortex cores play 
the role of normal region in the sense that gap drops to zero at the vortex 
axes. In Fig. \ref{doseclean}b) DC at $T=0.1T_c$, calculated according
to Eq. (\ref{dc}), is presented. At this temperature DC is thermally broadened
LDOS, but the oscillating pattern is still visible. 

\begin{figure}[t]
\includegraphics[angle=0,scale=0.45]{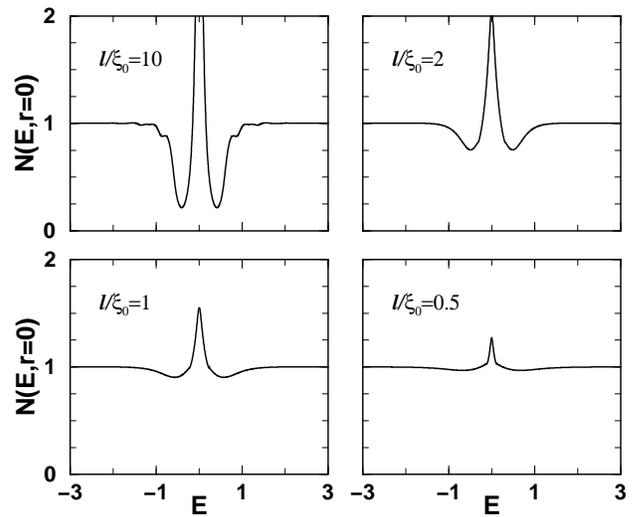}
\caption{LDOS at the vortex center as a 
function of energy plotted for various values of mean free path $\ell$.
Normalized DC $\sigma/\sigma_N$ is almost indistinguishable from 
the normalized LDOS $N(E,\bm r=0)/N_0$ at this temperature $T=0.1T_c$.}
\label{dosedirty}
\end{figure}

The coherency of quasiparticles is essential both for zero-bias peak and
oscillation of LDOS with energy at the vortex axis. In Fig. {\ref{dosedirty}} 
LDOS  at the vortex center as a function of energy is 
plotted for various values of impurity concentration. Oscillation amplitude is 
very sensitive to the presence of impurities and is almost lost even in very 
clean samples with $\xi_0/\ell=0.1$. Proliferating impurity concentration will 
manifest as a flattening of LDOS at the vortex center: disappearance of 
zero-energy peak of LDOS, as well as disappearance of deep minima for
$E<\Psi(B=0)$. In the dirty limit $\xi_0/\ell\rightarrow \infty$, LDOS at the 
vortex center is equal to $N_0$ for all quasiparticle energies 
$N(\bm r=0,E)=N_0$. \cite{wats,golubov} 

\section{Local density of states  and specific heat}
\label{ldosandsh}

The low energy quasiparticle excitations play the important role 
in the low temperature thermodynamics. Specific heat $C_s(T)$
of a superconductor is given by

$$
\frac{C_s}{T}=
2\int\limits_{-\infty}^\infty
\mathrm{d}E\,\frac{\partial \overline{N(E)}}{\partial T}
\left\{
\ln\left[
2\cosh\left(
\frac{E}{2T}
\right)
\right]-\right.
$$
\begin{equation}
\left.
\frac{E}{2T}\tanh\frac{E}{2T}
\right\}+
2\int\limits_{-\infty}^\infty
\frac{E^2}{4T^3}
\frac{\overline{N(E)}{\rm d}E}{\cosh^2\left[
\displaystyle
\frac{E}{2T}
\right]}.
\end{equation}
One can utilize this expression only if the energy
dependent, spatially averaged, LDOS $\overline{N(E)}$ is
provided. However, in the limit $T\rightarrow 0$ the
first integral is zero. 
For small $T$ the function to be integrated in the second
integral is nonzero only in the small vicinity of $E=0$.
Therefore we can replace $\overline{N(E)}$ 
by $\overline{N(E=0)}$
\begin{equation}
\lim\limits_{T\rightarrow 0}
\frac{C_s}{T}=2\int\limits_{-\infty}^\infty
\frac{E^2}{4T^3}
\frac{\overline{N(E=0)}\mathrm{d}E}{\cosh^2\left[
\displaystyle
\frac{E}{2T}
\right]}=
\frac{2\pi^2\overline{N(E=0)}}{3}.
\end{equation}
In the normal phase $C_n/T=2\pi^2 N_0/3$
which gives us the well known result 
\begin{equation}
\lim\limits_{T\rightarrow 0}
\frac{C_s}{C_n}=\frac{\overline{N(E=0)}}{N_0}\,.
\label{ne}
\end{equation}

\begin{figure}[t]
\vbox{
\hspace{-1.5cm}
\includegraphics[angle=270,scale=0.35]{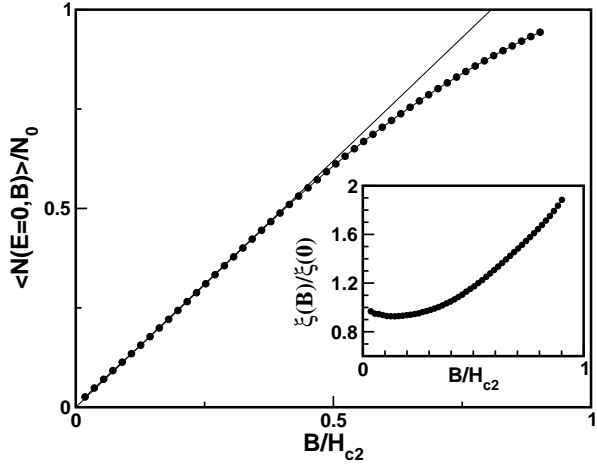}}
\vspace{-1cm}
\caption{Field dependence of spatially averaged zero-energy
LDOS in the clean limit. Straight line is guide for eye. In the inset is
shown normalized core size $\xi(B)/\xi(0)$ as a function of $B/H_{c2}$.}
\label{dosclean}
\end{figure}

If the low energy quasiparticles are localized in the 
vortex cores, which is true for $s$-wave superconductors
at least in the limit of very small fields, then 
$\overline{N(E=0)}\sim \rho^2/S_{\rm cell}$.
Here $S_{\rm cell}=\Phi_0/B$ is vortex lattice cell area
and $\rho$ is the size of the vortex core. If we further
assume that $\rho^2\sim \Phi_0/H_{c2}$ then we arrive at 
the following scaling relationship $\overline{N(E=0)}
\sim B/H_{c2}$, for $s$-wave superconductors. 
However there is a number of reports of nonlinear
field dependence of $\gamma_s(H)$ in $s$-wave
superconductors. One of the offered explanations is
that vortex core size $\rho$ itself is field dependent
which in turn lead to the nonlinear field dependence of
zero-energy DOS. The shrinking of the vortex core
with increasing field is detected in NbSe$_2$\cite{sonnier1}
and YBa$_2$Cu$_3$O$_{6.60}$. \cite{sonnier2} This is further supported 
by numerical calculations in dirty\cite{golubov,sonnier1} 
and clean \cite{ichioka} limit. Such an explanation 
brings out another puzzle. Experimental
study on influence of nonmagnetic impurities on the 
$\gamma(H)$ in 
Y$\left(\right.$Ni$_{1-x}$Pt$\left._x\right)_2$B$_2$C and 
Nb$_{1-x}$Ta$_x$Se$_2$ revealed that linear
$\gamma(H)$ is achieved only in dirty samples.\cite{nohara} This 
result suggest that the vortex core size in the dirty 
superconductors is field independent. Numerical calculation
by Golubov and Hartman,\cite{golubov} as well as Sonnier 
{\it et al.}\cite{sonnier1} shows
quite contrary, that even in the dirty limit $\rho$ should
shrink with increasing field.

Here we emphasize the necessity to perform the calculation at low temperature
in order to analyze the specific heat data through the ZEDOS.
In Ref. \onlinecite{ichioka} calculation performed at $T=0.5T_c$ revealed 
that $\overline{N(E=0)}\sim\xi^2(B)B$, where $\xi(B)$ is independently 
calculated vortex core radius. At lower temperatures, due to the Kramer-Pesch
effect, the core radius is smaller and it might have different field 
dependence.   

The result for the field dependence of ZEDOS in the clean limit,
for $T=0.1T_c$, is shown in Fig \ref{dosclean}. 
In the inset we plot the field dependence
of the core radius at the same temperature. We define the core radius
$\xi$ as $1/\xi=(\partial |\Psi(r)|/\partial r)/|\Psi_{NN}|$ 
where $|\Psi_{NN}|$ is the maximum value of the order parameter along the 
nearest neighbor direction, and derivative is taken along the same direction.
Compared to the previously reported result at higher temperature 
$T=0.5T_c$, where $\xi(B)$ decreases with field,\cite{ichioka} 
at $T=0.1T_c$, vortex core radius is rather constant at low fields. 
As a consequence, zero-energy LDOS is also linear function of magnetic 
induction.

\begin{figure}[t]
\includegraphics[angle=0,scale=0.35]{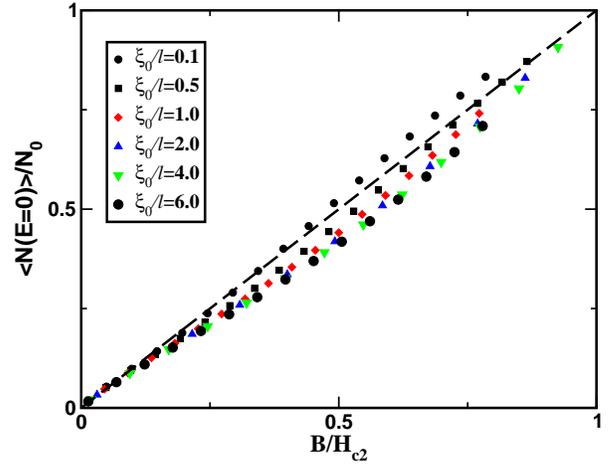}
\caption{Field dependence of spatially averaged zero-energy
LDOS for various mean free path.}
\label{shdirty}
\end{figure}

In the clean limit ZEDOS in between vortexes is negligible in
fields as large as $B=0.4H_{c2}$. In other words, the main contributions to 
ZEDOS is coming from the vortex cores. On the other hand, in the dirty limit,
ZEDOS is not confined to the vortex cores, but it is spread throughout the 
vortex lattice cell. It is large even in between vortexes. Thus, the scaling
relation $\overline{N(E=0)}\sim\xi^2(B)B$ is of no use in the dirty limit.
This is the reason why we do not attempt to correlate vortex core size
$\xi(B)$ and field dependence of LDOS in the impure case. However, 
$\overline{N(E=0,B)}$ is a linear function of magnetic induction at low 
fields for any impurity concentration: $\overline{N(E=0,B)}/N_0=
c(\tau)B/H_{c2}$. Constant of proportionally $c(\tau)$ weakly depends
on the electron mean free path and saturates to $c\approx 0.8$ in the dirty 
limit. Numerical calculation of  $\overline{N(E=0,B)}/N_0$ as a function
of mean-free path value is presented in Fig. \ref{shdirty}. We note the 
concave curves for dirtier cases. This behaviors coincide with the analysis 
near $H_{c2}$ by Kita.\cite{kita}

\begin{figure}[t]
\includegraphics[angle=0,scale=0.35]{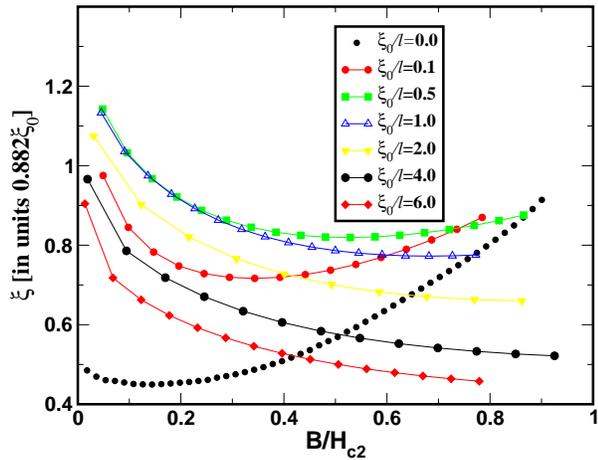}
\caption{Field dependence of vortex core size  
for various mean free path.}
\label{xidirty}
\end{figure}

In Fig. \ref{xidirty} is shown the field dependence of the core radius as 
calculated from the
pair potential profile $\Psi(\bm r)$. For a fixed relative field
$B/H_{c2}$ core radius $\xi$ is a nomonotonic function of mean free
path, first sharply increases and then slowly decreases with 
increasing of ratio $\xi_0/\ell$. In the dirty limit vortex core
shrinks with increasing field, which is consistent with the previous
calculations,\cite{golubov,sonnier1} in sharp contrast with
vortex core enlargement with increasing field in the clean limit. 

The experimental data, however, revealed that constant $c=1$ in the 
dirty limit.\cite{nohara} It also shows that scaling $\overline{N(E=0,B)}/N_0=
c(\tau)B/H_{c2}$ for all field values, which is a remarkable feature 
that still lacks the explanation. Worth is mentioning that in Ref.
\onlinecite{lipp} specific heat is a nonlinear function 
of field in samples Y$\left(\right.$Ni$_{1-x}$Pt$\left._x\right)_2$B$_2$C
for all $0<x<1$.  In these materials, we need to consider also the effect 
of gap anisotropy.

\section{Summary}

In this paper we examined the effect of impurities on LDOS
in isotropic $s$-wave superconductors. We showed that coherency
is crucial in forming the spatial structure of LDOS. As soon as
impurities are introduced into the superconductor, scattered 
electrons loose the information on their initial state, the coherency
is lost and sharp LDOS structure is flattened. It is calculated how
DC spectra evolve with electron mean-free path. Although the impurities
have a great impact on LDOS, spatially averaged LDOS shows weak dependence
on relative field $B/H_{c2}$. We hope that present calculation can be helpful
to roughly estimate the electron mean free path through the LDOS
measurement.

\acknowledgments

We acknowledge the useful communication with T. Dahm at the initial stage of 
this work.


\begin{thebibliography}{99}

\bibitem{hess} H. F. Hess, R. B. Robinson, R. C. Dynes, J. M. Valles, Jr., 
and J. V. Waszczak,
Phys. Rev. Lett. {\bf 62}, 214 (1989). 

\bibitem{hess1} H. F. Hess, R. B. Robinson, and J. V. Waszczak,
Phys. Rev. Lett. {\bf 64}, 2711 (1990). 


\bibitem{nohara} 
M. Nohara, M. Isshiki, F. Sakai, and H. Takagi, 
J. Phys. Soc. Jpn. {\bf 68}, 1078 (1999).

\bibitem{lipp} D. Lipp, M. Schneider, A. Gladun, S.-L. Drechsler, 
J. Freudenberger, G. Fuchs, K. Nenkov, K.-H. M\" ulcer, T. Cichorek, and 
P. Gegenwart, Europhys. Lett. {\bf 58}, 435 (2002). 

\bibitem{renner} Ch. Renner, A. D. Kent, Ph. Niedermann, and \O. Fischer,
Phys. Rev. Lett. {\bf 67}, 1650 (1991).

\bibitem{ullah} S. Ullah, A. T. Dorsey, and L. J. Buchholtz,
Phys. Rev. {\bf B} 42, 9950 (1990).

\bibitem{kleinsingle} U. Klein, Phys. Rev. {\bf B} 41, 4819 (1990).


\bibitem{ichioka} M. Ichioka, N. Hayashi, and K. Machida,
Phys. Rev. B {\bf 55}, 6565 (1997).


\bibitem{eschrig}M. Eschrig: Ph. D Thesis, University of Bayreuth (1997);
cond-mat/9804330.

\bibitem{kato}Y. Kato, and N. Hayashi, 
J. Phys. Soc. Jpn. {\bf 71}, 1721 (2002).


\bibitem{dukan} S. Dukan, and Z. Te\v sanovi\' c, 
Phys. Rev. {\bf B} 56, 838 (1997).

\bibitem{wats} R. Watts-Tobin, L. Kramer, and W. Pesch, 
J. Low Temp. Phys. {\bf 17}, 71 (1974).

\bibitem{golubov} A. A. Golubov, and U. Hartmann, 
Phys. Rev. Lett. {\bf 72}, 3602 (1994).

\bibitem{lowk} P. Miranovi\' c, and K. Machida,
Phys. Rev. B {\bf 67}, 092506 (2003).

\bibitem{klein} U. Klein, J. Low Temp. Phys. {\bf 69}, 1 (1987).

\bibitem{thuneberg} E. V. Thuneberg, J. Kurkij\" avri, and
D. Rainer, Phys. Rev. B {\bf 29}, 3913 (1984).

\bibitem{remark} Numerical solution of Bogoliubov-deGennes
equations revealed a little difference between LDOS in
two-dimensional and three-dimensional case, see
J. D. Shore, M. Huang, A. T. Dorsey, and J. P. Sethna,
Phys. Rev. Lett. {\bf 62}, 3089 (1989).


\bibitem{schopohl} N. Schopohl, cond-mat/9804064.

\bibitem{schopohlandmaki} N. Schopohl, and K. Maki, 
Phys. Rev. B {\bf 52}, 492 (1995).


\bibitem{klein1} U. Klein, Phys. Rev. B {\bf 41}, 4819 (1990).

\bibitem{dahm} T. Dahm, S. Graser, C. Iniotakis, and N. Schophol,
Phys. Rev. B {\bf 66}, 144515 (2002).

\bibitem{rainer} D. Rainer, J. A. Sauls, and D. Waxman, 
Phys. Rev. B {\bf 54}, 10094 (1996).


\bibitem{hayashi} N. Hayashi, M. Ichioka, and K. Machida, 
Phys. Rev. B {\bf 56}, 9052 (1997). 


\bibitem{pottinger}B. P\"ottinger, and U. Klein,
Phys. Rev. Lett. {\bf 70}, 2806 (1993).

\bibitem{degennes}P. G. De Gennes, and D. Saint-James,
Phys. Lett. {\bf 4}, 151 (1963).

\bibitem{tomasch} W. J. Tomasch, 
Phys. Rev. Lett. {\bf 15}, 672 (1965).

\bibitem{rowell} J. M. Rowell, and W. L. McMillan,
Phys. Rev. Lett. {\bf 16}, 453 (1966).

\bibitem{sonnier1}
J. E. Sonier, R. F. Kiefl, J. H. Brewer, J. Chakhalian, S. R. Dunsiger, 
W. A. MacFarlane, R. I. Miller, A. Wong, G. M. Luke, and J. W. Brill, 
Phys. Rev. Lett. {\bf 79}, 1742 (1997).

\bibitem{sonnier2}  J. E. Sonier, J. H. Brewer, R. F. Kiefl, D. A. Bonn, 
S. R. Dunsiger, W. N. Hardy, R. Liang, W. A. MacFarlane, R. I. Miller,  
T. M. Riseman, D. R. Noakes, C. E. Stronach, and M. F. White, Jr.,
Phys. Rev. Lett. {\bf  79}, 2875 (1997). 

\bibitem{kita} T. Kita, private communication.
 
\end{thebibliography}
\end{document}